# A LARGE LUNAR IMPACT BLAST ON SEPTEMBER 11[TH] 2013

**José M. Madiedo[1, 2], José L. Ortiz[3], Nicolás Morales[3] and Jesús Cabrera-Caño[2]**

1 Facultad de Física, Universidad de Sevilla, Departamento de Física Atómica, Molecular y Nuclear, 41012 Sevilla, Spain.
[2] Facultad de Ciencias Experimentales, Universidad de Huelva. 21071 Huelva (Spain).
[3] Instituto de Astrofísica de Andalucía, CSIC, Apt. 3004, Camino Bajo de Huetor 50, 18080 Granada, Spain.

**ABSTRACT**

On 2013 September 11 at 20h07m28.68 ± 0.01 s UTC, two telescopes operated in the framework of our lunar impact flashes monitoring project recorded an extraordinary flash produced by the impact on the Moon of a large meteoroid at selenographic coordinates 17.2 ± 0.2 º S, 20.5 ± 0.2 º W. The peak brightness of this flash reached 2.9 ± 0.2 mag in V and it lasted over 8 seconds. The estimated energy released during the impact of the meteoroid was 15.6 ± 2.5 tons of TNT under the assumption of a luminous efficiency of 0.002. This event, which is the longest and brightest confirmed impact flash recorded on the Moon thus far, is analyzed here. The likely origin of the impactor is discussed. Considerations in relation to the impact flux on Earth are also made.

# 1 INTRODUCTION

The identification and analysis of flashes produced by the impact of meteoroids on the lunar surface is one of the techniques suitable for the study of the flux of interplanetary matter impacting the Earth. Hypervelocity impacts of projectiles on all sorts of targets generate optical radiation, the so called "flash" from the high temperature vaporized plasma. For lunar impact flashes it has been hypothesized that the radiation is also emitted from the condensing ejecta that cools down and form silicate droplets (Yanagisawa and Kisaichi 2002, Bouley et al 2012). The thermal emission from these droplets would cause longer-lasting flashes than those from the plasma. The first systematic attempts to identify impact flashes produced by large meteoroids striking the Moon by means of telescopic observations with CCD cameras date back to 1997 (Ortiz et al. 1999), but no conclusive evidence of impact flashes was recorded in that work. After that, impact flashes have been unambiguously detected during the maximum activity period of several major meteor showers by using this technique (e.g. Ortiz et al. 2000, Yanagisawa and Kisaichi 2002, Cudnik et al. 2002; Ortiz et al. 2002, Yanagisawa et al. 2006, Cooke et al. 2006), and flashes of sporadic origin have been also recorded (Ortiz et al. 2006, Suggs et al. 2008). This method of observing lunar flashes has the advantage over terrestrial meteor networks that the area covered by one single detection instrument is much larger than the atmospheric volume monitored by meteor detectors employed by fireball networks. The technique, which implies the systematic monitoring of the night side of the Moon, can be employed when the illuminated fraction of the lunar disk varies between, approximately, 5 and 60 %, i.e., during the first and last quarters. Besides, at least two telescopes must operate in parallel imaging the same area on the Moon in order to discard false detections produced by other phenomena such as, for instance, cosmic rays and electric noise. In addition, glints from artificial satellites and space debris can be confused with impact flashes if suitable fast imaging devices are not used.

Since 2009 our team is running a project named MIDAS, which is the acronym for Moon Impacts Detection and Analysis System. Its aim is to record and study impact flashes produced by the collision of meteoroids on the lunar surface by means of small telescopes. Previous observations of flashes produced by the collision of meteoroids of sporadic origin on the Moon's surface (Ortiz et al. 2006) indicate that the flux of materials impacting our planet would be higher than the flux predicted by Brown et al. (2002) from the analysis of fireballs in the atmosphere. So, additional observations of lunar impact flashes are desirable in order to analyze the reasons for such differences.

In this context, our systems recorded an extraordinary flash with a magnitude of 2.9 produced by the impact of a meteoroid on the lunar surface on 2013 September 11. With a duration of over 8 seconds, this is the brightest and longest confirmed impact flash ever recorded on the Moon.

Among the recorded 1999 Leonid impact flashes on the Moon two of them were considerably bright, of around magnitude 3 in the visible (e.g. Dunham 1999, Ortiz et al. 2000), although Cudnik et al. (2002) gave somewhat fainter magnitudes for the same events. Another bright Leonid in 1999 was reported by Yanagisawa and Kisaichi (2002). Its magnitude was brighter than 5 because at this level the detector saturated, but in this case the duration of the flash was longer (around 0.2 seconds) than the rest of the Leonids. This duration is still very short compared to the event that we report here. In 1953 a bright flash on the Moon was serendipitously registered in a photographic plate by Stuart (1953) while testing small telescopic equipment. Because the flash was not confirmed by any other instrument and because of the amateur observation, the real nature of the flash was not clear for many years and this event became another of the mysterious and often discredited Transient Lunar Phenomena. Nowadays, after the unambiguous observation of lunar impact flashes, it seems likely that the Stuart (1953) event was a real impact flash. Something similar happens with the Kolovos et al. (1988) flash caught in photography, whose cause was attributed to lunar outgassing by the authors at that time. Both the Stuart and Kolovos et al flashes now seem compatible with the phenomenology we have seen in lunar impact flashes in terms of brightness and duration. Here we analyze the September $11^{th}$ 2013 event, show its lightcurve and discuss several of its implications, including implications for the Earth impact hazard.

## 2 INSTRUMENTATION AND METHODS

The impact flash discussed here was imaged by our telescopes operating at our observatory in Sevilla, in the south of Spain (latitude: 37.34611 ºN, longitude: 5.98055 ºW, height: 18 m above the sea level). Our impact flashes monitoring system at this site employs two identical 0.36 m Schmidt-Cassegrain telescopes that image the same area of the Moon, but also a smaller Schmidt-Cassegrain telescope with a diameter of 0.28 m is available. All of them are manufactured by Celestron. These telescopes are endowed with monochrome high-sensitivity CCD video cameras (model 902H Ultimate, manufactured by Watec Corporation) which employ a Sony ICX439ALL 1/2" monochrome CCD sensor and produce interlaced analogue imagery according to the PAL video standard. Thus, images are obtained with a resolution of 720x576 pixels and a frame rate of 25 frames per second (fps). GPS time inserters are used to stamp time information on every video frame with an accuracy of 0.01 seconds. Besides, f/3.3 focal reducers manufactured by Meade are also employed in order to increase the area monitored by these devices. To maximize the monitored area on the Moon surface, each camera is oriented in such a way that the lunar equator is perpendicular to the longest side of the CCD sensor. Under these conditions, lunar features are easily identified in the earthshine and, so, these can be used to determine the selenographic coordinates (i.e., latitude and longitude on the lunar surface) of impact flashes.

When no major meteor showers are active, as it happened during the observing session where the impact discussed here was registered, our telescopes are oriented to an arbitrary region on the Moon surface in order to cover a common maximum area. The analogue video imagery generated by the cameras are continuously digitized and recorded on multimedia hard disks. Of course, the terminator is avoided in order to prevent saturation of the CCD sensors and also to avoid an excess of light from the illuminated side of the Moon in the telescopes. Even though the telescopes are tracked at nonsidereal lunar rates, recentering of the telescope is done manually from time to time because perfect tracking of the Moon at the required precision is not feasible with this equipment.

Once the observing session is over, the video streaming generated by each telescope was analyzed with the MIDAS software (Madiedo et al. 2010, 2011), which received the same name as our lunar impact flashes monitoring project but, when applied to this tool, is the acronym for Moon Impacts Detection and Analysis Software. This tool was developed to process live video streaming or AVI video files containing images of the night side of the Moon to automatically identify flashes produced by the impact of meteoroids on the lunar surface. In order to indentify an impact flash, the software compares consecutive video frames and detects brightness changes that exceed a given (user defined) threshold value (Madiedo et al. 2011). Then, if an event is detected, the software automatically provides its (x,y) coordinates on the image, but also the corresponding values of latitude and longitude on the lunar surface. These coordinates are those corresponding to the centroid of the flash. The same software is employed to perform the photometric analysis of these events.

### 3 OBSERVATIONS

On 2013 September 11, with a 6 day-old Moon, one of our 0.36m telescopes and the 0.28 m telescope were aimed at the same region of the night side of the lunar surface (Figure 1). The area monitored during that observing session by the CCD video devices attached to these telescopes, which was calculated with the MIDAS software, was of about $6.6 \cdot 10^6$ and $8.6 \cdot 10^6$ km$^2$, respectively. These cameras imaged an extraordinary flash on the lunar surface at 20h07m28.68 ± 0.01 s UTC (Figures 2 and 3). The event, which lasted about 8.3 seconds, had a peak visual magnitude of 2.9 ± 0.2. The calibration was determined as explained in the next paragraph. The recordings from both instruments confirmed that the flash was produced by the impact of a meteoroid, since it was simultaneously imaged at the same selenographic coordinates by both telescopes, and the centroid of the flash did not experience any relative motion with respect to that position during such time span, discarding satellite or space debris glints. Thus, according to the analysis performed with the MIDAS software, the impactor stroke the lunar surface at the coordinates 17.2 ± 0.2 ° S, 20.5 ± 0.2 ° W, which corresponds to the west part of Mare Nubium. The main circumstances of this impact are shown in Table 1.

The photometric analysis of the flash was performed with the MIDAS software and the result was double checked with the Limovie software (Miyashita et al. 2006). In a first step, we obtained the flash brightness expressed in device units (pixel value). The analysis was performed on a 28x28 pixels box around the flash. The same procedure is employed for reference stars, whose visual magnitude is known. Thus, by comparing the result obtained for the reference stars with that of the flash, the visual magnitude of the impact flash is inferred. The following stars in the Tycho-2 catalogue were considered: TYC1310-2697-1 (V magnitude 2.96), TYC6211-510-1 (V magnitude 4.46), TYC6152-832-1 (V magnitude 8.03), TYC5559-476-1 (V magnitude 7.50) and TYC5540-1438-1 (V magnitude 5.93). The lightcurve of the flash is shown in Figure 4. As can be noticed, there is a very rapid decrease of luminosity, so that the flash magnitude increases from 2.9 to 8, around a 5-magnitude decay, in about 0.25 seconds. This brightness decay rate is similar to that shown in Yanagisawa and Kisaichi (2002) for its brightest flash and also similar to the decay seen in the Ortiz et al. (2002) lightcurve of the brightest 2001 Leonid flash although in this latter case the decay was not smooth and seems somewhat longer. The total duration of the impact flash shown here is the longest ever observed because after the main decay the flux drops more smoothly till it reaches that background level in about 8 seconds. Figure 5 shows a sequence of images of the flash at different times zoomed in the impact area. The long duration of the flash reported here is, given its high brightness, consistent with the correlation between impact brightness and duration shown by Bouley et al. (2012).

## 4 RESULTS AND DISCUSSION

### 4.1. Impact Energy

The observed luminosity of the flash has been employed to determine the radiated power P, in Watts, from the following equation:

$$P = 3.75 \cdot 10^{-8} \cdot 10^{(-m/2.5)} f\pi\Delta\lambda R^2 \tag{1}$$

where m is the magnitude of the flash, $\Delta\lambda$ is the width of the filter passband (about 5000 Å), R is the Earth-Moon distance at the instant of the meteoroid impact (365300 km) and f is a factor that describes the degree of anisotropy of light emission. In the equation, $3.75 \ 10^{-8}$ is the flux density in W m$^{-2}$ μm$^{-1}$ for a magnitude 0 source according to the values given in Bessel (1979). For events where light is isotropically emitted from the surface of the Moon f=2, while f=4 if light is emitted from a very high altitude above the lunar surface. For the flash discussed here we have considered f=2 because we noticed that no surface features are illuminated by the flash so it cannot be very high above the lunar surface.

By numerically integrating this radiated power with respect to time, the energy released as visible light on the Moon ($E_r$) can be calculated. This magnitude is related to the kinetic energy E of the impactor by means of the following relationship:

$$E=\eta E_r \qquad\qquad\qquad (2)$$

where $\eta$ is the luminous efficiency (i.e. the fraction of the kinetic energy that is emitted in the visible). For this parameter we have assumed $\eta=2\cdot10^{-3}$ (the value determined for the Leonid lunar impact flashes in e.g. Bellot Rubio et al 2000, Ortiz et al. 2002) and was also used by Ortiz et al. (2006) to determine impact fluxes on Earth. This value is close to the $\eta=1.5\cdot10^{-3}$ value used by other investigators (Swift et al. 2011, Bouley et al. 2012). According to this, the kinetic energy of the impactor yields E=(6.5 ± 1.0)·$10^{10}$ J (15.6 ± 2.5 tons of TNT). Using the lower luminous efficiency by Swift et al. (2011) and Bouley et al. (2012) the resulting impact energy would be even higher than our estimation.

### 4.2. Impactor mass and source

On Earth, the association of a meteoroid with a given meteoroid stream is straightforward when the tracks of meteors produced by the ablation in the atmosphere of these particles of interplanetary matter are recorded. Thus, provided that the event is simultaneously detected from, at least, two different meteor observing stations, the radiant can be easily determined and the orbit of the meteoroid in the Solar System can be calculated (Ceplecha 1987). For the calculation of this orbit the knowledge of the velocity vector is fundamental. Once radiant and orbital data are available, the meteoroid can be associated with a given meteoroid stream. However, for meteoroid impacts taking place on the lunar surface the velocity vector is unknown, since just the impact position is available from observations. So, the above mentioned approach cannot be employed and, in fact, in this case it is not possible to unambiguously associate an impact flash with a given meteoroid stream. Nevertheless, since no major meteor shower was active by the time of the detection of the impact flash discussed here (Jenniskens 2006), the event could be considered, in principle, as the result of the collision of a sporadic meteoroid. In this case, the average impact velocity V on the lunar surface would be of about 17 km s$^{-1}$ (Ortiz et al. 1999). The impactor mass M has been obtained from the kinetic energy of the meteoroid (E):

$$M=2EV^{-2} \tag{3}$$

According to this, the meteoroid mass yields M=450 $\pm$ 75 kg. To calculate the meteoroid size we have considered a bulk density ranging between 0.3 g cm$^{-3}$ (the corresponding to soft cometary materials) to 3.7 g cm$^{-3}$ (the corresponding to ordinary chondrites) (Ceplecha 1988). Thus, the diameter of the meteoroid would range between 142 $\pm$ 9 and 61 $\pm$ 3 cm, respectively.

However, on 2013 September 9, two days prior to the lunar impact flash between 21h30m and 23h20m UTC, a very minor meteor shower, the September ε-Perseid meteor shower (SPE) exhibited an outburst of its activity, producing a display of bright meteors (most of them ranging between magnitude 4 and -8). This outburst peaked around 22h22m UTC, with a rate of about 3 meteors per minute (Jenniskens 2013). Although the rate of this shower was back to normal after September 10d8h UTC, it remained active during the following days, since its activity period extends up to about September 23 (Jenniskens 2006). When the SPE were taken into consideration, we obtained with the MIDAS software that the impact flash discussed here was compatible with the impact geometry of meteoroids belonging to this stream (Figure 1), which opens the possibility that the particle was not a sporadic. With a geocentric velocity of about 64.5 km s$^{-1}$ (Jenniskens 2006), SPE meteoroids would impact the Moon with a velocity which is considerably higher than the average impact velocity of sporadic meteoroids. However, it must be taken into account that this geocentric velocity must be corrected to find the correct impact velocity on the lunar surface. Thus, a correcting factor for the kinetic energy has to be applied, since the gravitational field of our planet gives rise to a larger impact velocity on Earth compared to the lunar case. For sporadic meteoroids, which can impact from random directions, this factor is around 1.4 (Ortiz et al. 2006). For meteoroids belonging to the September ε-Perseid stream, we have found that the impact velocity is of about 53.2 km s$^{-1}$, which means that in this case this factor is 1.2. This impact velocity has been obtained from a straightforward computation of the relative velocity of SPE meteoroids with respect to the Moon from the known values of the heliocentric velocity vector of the Moon obtained from the JPL Horizons online ephemeris system (http://ssd.jpl.nasa.gov/horizons.cgi), the heliocentric velocity vector of Earth (obtained from the same source) and the known geocentric velocity of SPE meteoroids. Thus, by following the above-described approach, the impactor mass would be much lower in this case, of about $46 \pm 7$ kg. Besides, by using an average bulk density for cometary meteoroids of 1.8 g cm$^{-3}$ (Babadzhanov and Kokhirova 2009), the meteoroid diameter yields $36 \pm 2$ cm. However, according to equations (1) and (2) in (Hughes 1987), SPE meteoroids producing mag -8 fireballs (the brightest SPE bolides recorded during the outburst according to Jenniskens (2013)), would have a mass or around 70 g. So, given that the size of the impactor is considerably higher than the largest meteoroids that caused the outburst of the SPE stream, and given that this outburst was more than one day earlier than our impact flash, we tend to think that a sporadic origin is perhaps more likely.

## 4.2. Crater size

To estimate the size of the crater produced by the impact of the meteoroid we have employed the following crater-scaling equation (Schmidt and Housen 1987, Melosh 1989):

$$D = \gamma^{-0.26} M^{0.26} V^{0.44} \tag{4}$$

where

$$\gamma = 0.31 g^{0.84} \rho_p^{-0.26} \rho_t^{1.26} \left( \sin 45 / \sin \theta \right)^{1.67} \tag{5}$$

In these relationships units are in the mks system. D is the crater diameter, M is the impactor mass and V its velocity, g is the gravitational acceleration, $\rho_p$ and $\rho_t$ are the impactor and target bulk densities, respectively, and $\theta$ is the impact angle with respect to the vertical. For the target bulk density we have taken $\rho_t$=2700 kg m$^{-3}$.

If the meteoroid is associated with a sporadic source, the impact angle $\theta$ is unknown. In this case, we have used for this parameter the value of the most likely impact angle: 45º. Then, from equations (4) and (5), the crater diameter yields D=47 m for an impactor bulk density of 0.3 g cm$^{-3}$ and D=56 m for $\rho_p$= 3.7 g cm$^{-3}$.

On the other hand, if the meteoroid belonged to the September ε-Perseid meteoroid stream the impact angle would be of about 39º with respect to the local vertical, according to the impact geometry shown in Figure 1. In this way, equations (4) and (5) yield D= 46 m for $\rho_p$= 1.8 g cm$^{-3}$.

The derived sizes are small for ground based observatories to identify them, but lunar orbiters can take images of the impact regions to recognize fresh craters and study them. The derived crater size is the largest for an impact flash ever reported and if the crater is identified, the measurement of its diameter would allow us to give further constraints on the luminous efficiency, which is a poorly characterized parameter and is important to refine the impact flux on Earth.

## 4.4. Implications for the terrestrial impact hazard

In the following discussion about impact rates on the Earth, we have used energy rather than mass, since expressing impact rates as a function of the impactor mass would require a correct choice for the impact velocity. However, for impact rates given as a function of impactor energy no critical assumptions about meteoroid velocity are necessary.

According to the kinetic energy inferred for the impact flash discussed here, the impact rate on the whole Moon for fragments with an energy above 15.6 tons of TNT would be of about 126 events per year, by considering the total observing time employed by our team since 2009 (around 300 hours) and the average lunar area monitored by our telescopes during that time span (about $8.8 \cdot 10^6$ km$^2$). This lunar impact rate can be translated into the corresponding terrestrial impact rate by scaling it according to the surface area of our planet (about 13.5 higher than that of the Moon) and by taking into account a 1.3 gravitational focusing factor for the flux (Ortiz et al. 2006). In addition, the previously mentioned correcting factor for the kinetic energy has to be also applied (1.4 if we assume that the impactor was a sporadic meteoroid and 1.2 if it belonged to the September ε-Perseid stream). Thus, the impact energy of the lunar impact flash would be equivalent to an impact energy of 28.3 ± 4.5 tons of TNT on Earth for the sporadic meteoroid, and $24.3 \pm 3.8$ tons of TNT for the SPE meteoroid. So, by performing the corresponding surface area scaling between both bodies, the impact rate on Earth for events with an energy above these values would be of about $1680 \pm 1050$ events per year (Figure 6). This is considerably higher than the ~90 events per year predicted for this impact energy by Brown et al. (2002), but is in agreement with the impact flux distribution obtained by Ortiz et al. (2006). In fact, Ortiz et al. (2006) showed that a luminous efficiency for impact flashes of about 0.02 would be necessary to provide results consistent with the terrestrial impact rate predicted by Brown et al. (2002), although such an efficiency would be incompatible with the observations of Leonid impact flashes on the Moon and with hypervelocity impact experiments. The analysis of a Perseid lunar impact flash seems also to be inconsistent with a 0.02 luminous efficiency because the size distribution of the Perseid meteoroid stream would have to be too steep (Yanagisawa et al. 2006). Thus, Ortiz et al. (2006) suggested that the impact hazard estimates given by Brown et al. (2002) were too low, and an enhancement of at least a factor 3 in the terrestrial impact rate would be necessary to match the results obtained from the observations of lunar impact flashes. Our analysis of the impact flash discussed in this work also supports this idea, as also do recent observations of superbolides over Spain (Madiedo et al. 2013) and is also consistent with other studies on the fluxes of fireballs (Ceplecha 2001) not used in the Brown et al. (2002) work. While the present paper was in review phase Brown et al. (2013) have revised their impact hazard calculations in Brown et al. (2002). Even though they do not provide an accurate figure of the upward increase that they found, they mention in the order of a factor 10 increase, which is coincident with our requirements.

## 5 CONCLUSIONS

We have analyzed the impact flash that took place on the Moon on 2013 September 11 at 20h07m28.68 ± 0.01 s UTC, during the waxing phase. The conclusions derived from this research are listed below:

1) With a peak brightness equivalent to mag. 2.9 ± 0.2 and a duration of 8.3 seconds, this is the brightest and longest confirmed impact flash ever recorded on the lunar surface. The energy released during the impact was of 15.6 ± 2.5 tons of TNT assuming a luminous efficiency of 0.002. The event occurred on the west part of Mare Nubium at coordinates 17.2 ± 0.2 ° S, 20.5 ± 0.2 ° W.

2) Two sources have been considered for the impactor. The event was compatible with the impact geometry of the September ε-Perseids minor shower, but it could also be associated with a sporadic meteoroid. By considering a luminous efficiency of $2·10^{-3}$, the impactor mass would be of about 450 kg for the sporadic meteoroid, and around 46 kg if the particle belonged to the SPE meteoroid stream.

3) The crater produced by this impact would be of about 46 m for a SPE meteoroid. For a sporadic event, this diameter would range

between 47 m (for a bulk density $\rho_p$=0.3 g cm$^{-3}$) and 56 m (for $\rho_p$ =3.7 g cm$^{-3}$). The identification of this crater in order to compare its actual size with the values obtained from our analysis could be a target for any current or future spacecraft orbiting the Moon. The actual size would help to constraint the luminous efficiency considerably, which is important for impact hazard computations.

4) This event exemplifies that Earth impact hazard estimations were not well constrained because we derive a value which is one order of magnitude above the estimates by Brown et al. (2002). While our paper was in review phase Brown et al. (2013) have reconsidered their original calculations with new data and now they report an increased impact hazard, although the exact factor is still uncertain. Thus, a systematic monitoring of moon impact flashes but also of fireballs in the Earth's atmosphere would provide a more reliable impact frequency, especially if the luminous efficiency is well calibrated.

**ACKNOWLEDGEMENTS**

The authors acknowledge support from Junta de Andalucía (project P09-FQM-4555). Support from AYA2011-30106-C02-01 and FEDER funds is also acknowledged

**REFERENCES**

Babadzhanov P.B. and Kokhirova G.I., 2009, A&A 495, 353.

Bessel M.S., 1979, Publications of the Astronomical Society of the Pacific 91, 589.

Bouley S. et al., 2012, Icarus, 218, 115.

Brown P., Spalding R.E., Revelle D.O., Tagliaferri E., Worden S.P, 2002, Nature, 420, 294.

Brown P. et al., 2013, Nature, 503, 238.

Ceplecha Z., 1987, Bull. Astron. Inst. Cz., 38, 222.

Ceplecha Z., 1988, Bull. Astron. Inst., 39, 221.

Ceplecha Z., 2001, In: Collisional processes in the solar system. Marov M.Y. and Rickman H. (Eds.), Astrophysics and space science library, Kluwer Academic Publishers, Vol. 261, p. 35.

Cooke W.J., Suggs R.M., Swift W.R., 2006, Lunar Planet. Sci. 37. Abstract 1731.

Cudnik B.M., Dunham D.W., Palmer D.M., Cook A.C., Venable J.R., Gural P.S., 2002, Lunar Planet. Sci. 33. Abstract 1329C.

Dunham D.W., 1999, IAU Circ 7320.

Hughes D.W., 1987, A&A, 187, 879.

Jenniskens P., 2006, Meteor Showers and their Parent Comets. Cambridge University Press.

Jenniskens P., 2013, Central Bureau Electronic Telegrams, 3652, 2.

Kolovos G., Seiradakis J.H., Varvoglis H., Avgoloupis S., 1988, Icarus, 76, 525.

Madiedo J.M., Trigo-Rodriguez J.M., Ortiz J.L., Morales N., 2010, Advances in Astronomy, doi:10.1155/2010/167494.

Madiedo J.M., Ortiz J.L., Morales N., 2011, EPSC-DPS Joint Meeting 2011, Abstract #Vol. 6, EPSC-DPS2011-66-0.

Madiedo J.M. et al, 2013, Icarus, in press.

Melosh H.J., 1989. Impact Cratering: A Geologic Process. Oxford Univ. Press, New York.

Miyashita K., Hayamizu T., Soma M., 2006, Report of the National Astronomical Observatory of Japan, 9, 1.

Ortiz J.L., Aceituno F.J., Aceituno J., 1999. A&A, 343, L57.

Ortiz J.L., Sada P.V., Bellot Rubio L.R., Aceituno F.V., Aceituno J., Gutierrez P.J., Thiele U., 2000, Nature, 405, 921.

Ortiz J.L., Quesada J.A., Aceituno J., Aceituno F.J., Bellot Rubio L.R., 2002, ApJ, 576, 567.

Ortiz J.L., Aceituno F.J., Santos-Sanz P., Quesada J.A., 2005, In: American Astronomical Society, DPS Meeting 37, 17.05.

Ortiz J.L. et al., 2006, Icarus, 184, 319.

Schmidt R.M, Housen K.M., 1987, Int. J. Impact Eng., 5, 543.

Stuart L.H., 1956, Strolling Astron., 10, 42.

Suggs R.M., Cooke W., Suggs R., McNamara H., Swift W., Moser D., Diekmann A., 2008, Bulletin of the American Astronomical Society, 40, 455.

Swift W.R., Moser D.E., Suggs R.M., Cooke W.J., 2011, In: Meteoroids: The Smallest Solar System Bodies, Edited by W.J. Cooke, D.E. Moser, B.F. Hardin, and D. Janches, NASA/CP-2011-216469, p 125.

Yanagisawa M., Kisaichi N., 2002, Icarus, 159, 31.

Yanagisawa M., Ohnishi K., Takamura Y., Masuda H., Ida M., Ishida M., 2006, Icarus, 182, 489.

# **TABLES**

| | |
|---|---|
| Date and time | 2013 Sept. 11 at 20h07m28.68±0.01s UTC |
| Peak brightness | 2.9±0.2 in visual magnitude |
| Selenographic coordinates | Lat.: 17.2±0.2 º S, Lon.: 20.5±0.2 º W |
| Duration (s) | 8.3 |
| Impact energy | $(6.5±1.0)·10^{10}$ J (15.6±2.5 tons of TNT) |
| Equivalent impact energy on Earth | SPO: $(1.2±0.2)·10^{11}$ J (28.3±4.5 tons of TNT) |
| | SPE: $(1.0±0.2)·10^{11}$ J (24.3±3.8 tons of TNT) |
| Meteoroid mass (kg) | SPO: 450±75 |
| | SPE: 46±7 |
| Meteoroid diameter (cm) | SPO: 142±9 ($\rho_p$=0.3 g cm$^{-3}$); 61±3 ($\rho_p$=3.7 g cm$^{-3}$) |
| | SPE: 36±2 ($\rho_p$=1.8 g cm$^{-3}$) |
| Meteoroid impact velocity (km s$^{-1}$) | SPO: 17 |
| | SPE: 53.2 |
| Impact angle (º) | SPO: 45 º |
| | SPE: 39 º |
| Crater diameter (m) | SPO: 47 ($\rho_p$=0.3 g cm$^{-3}$); 56 ($\rho_p$=3.7 g cm$^{-3}$) |
| | SPE: 46 ($\rho_p$=1.8 g cm$^{-3}$) |

Table 1. Observed data and some estimated values of the lunar impact flash discussed in this work, by assuming an impact efficiency η=2·10$^{-3}$. SPO indicates a meteoroid with a sporadic origin, while SPE indicates a meteoroid belonging to the September ε-Perseid stream.

## **FIGURES**

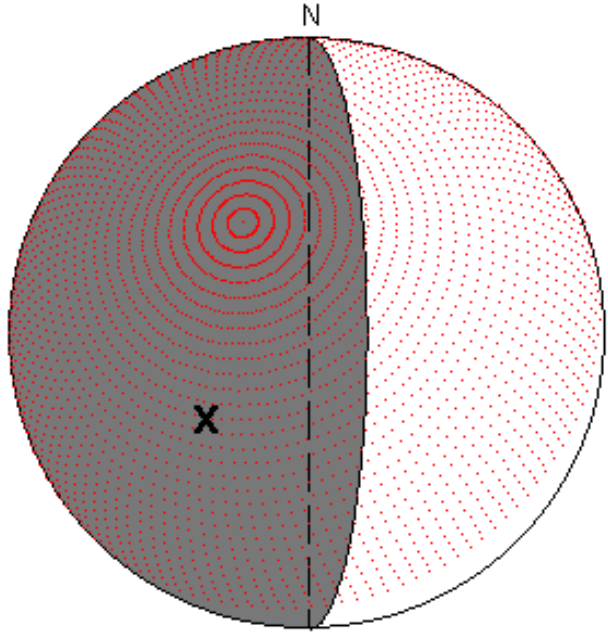

Figure 1. The lunar disk as seen from our planet on 2013 September 11. The gray region corresponds to the night side and the white region is the area illuminated by the Sun. The dotted region corresponds to the area where meteoroids from the September ε-Perseid stream could impact. The position of the impact flash discussed here is marked with an X.

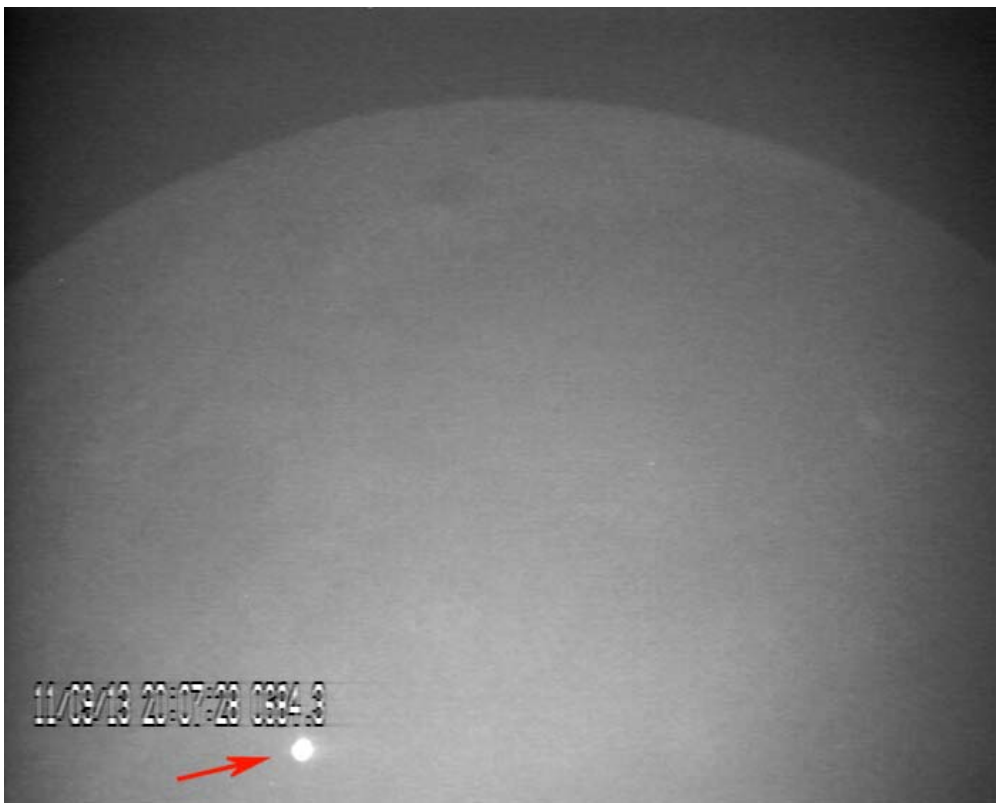

Figure 2. Impact flash detected from Sevilla by the 0.36 m telescope on
2013 September 11 at 20h07m28.68 ± 0.01 s UTC.

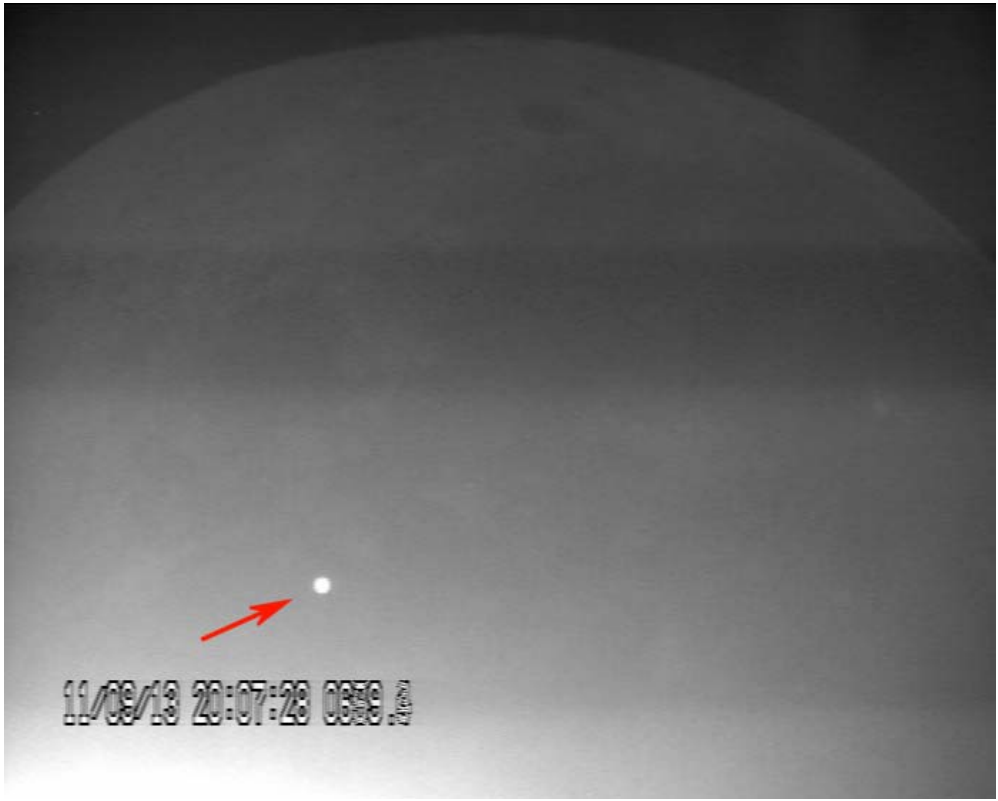

Figure 3. Impact flash detected from Sevilla by the 0.28 m telescope on
2013 September 11 at 20h07m28.68 ± 0.01 s UTC.

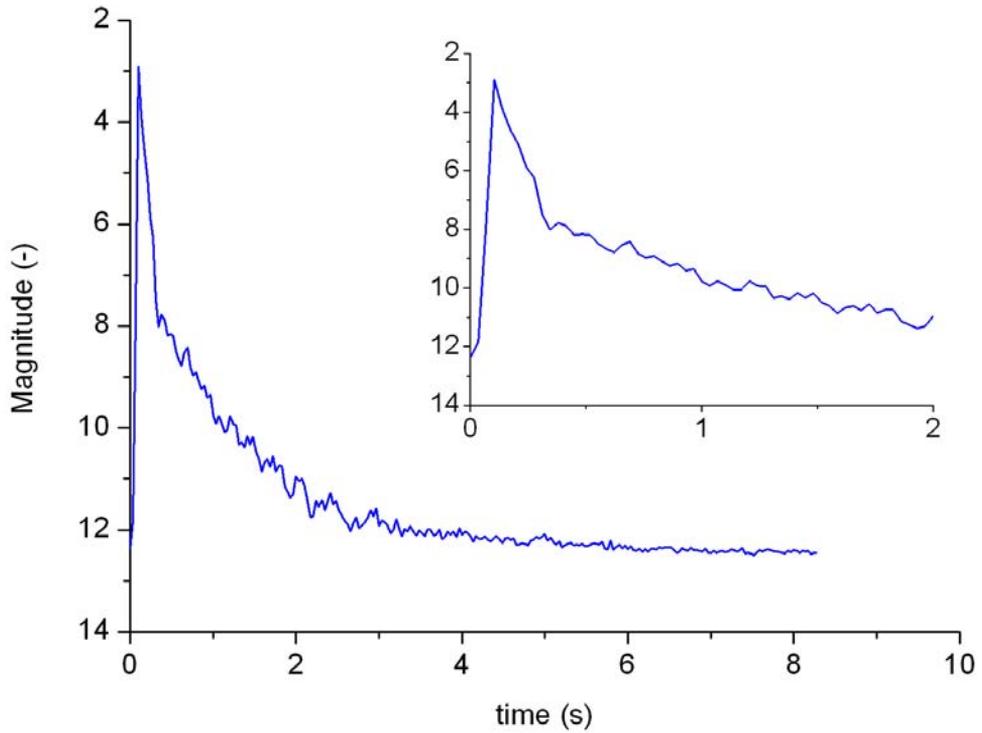

Figure 4. Lightcurve (V magnitude vs. time plot) obtained for the impact flash. The insert shows the evolution of magnitude during the first two seconds.

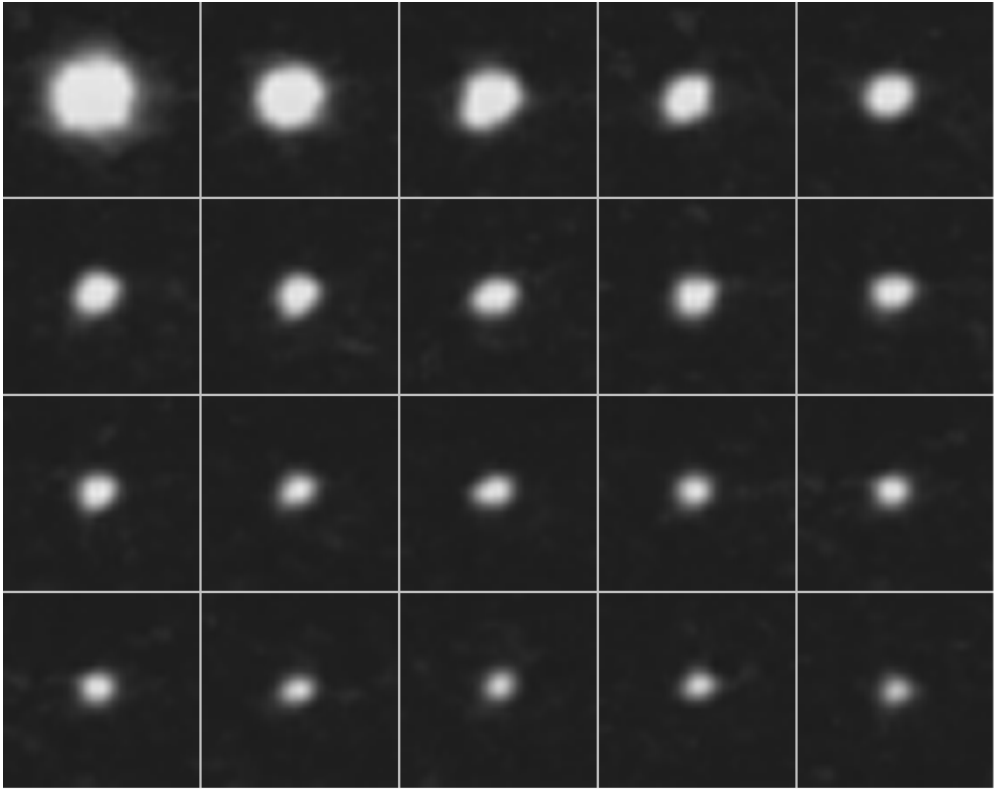

Figure 5. Mosaic of zoomed images showing the flash evolution with time during the first two seconds. Time increases from left to right on each row, starting from the upper left. The time interval between two consecutive images on the same row is 0.1 seconds

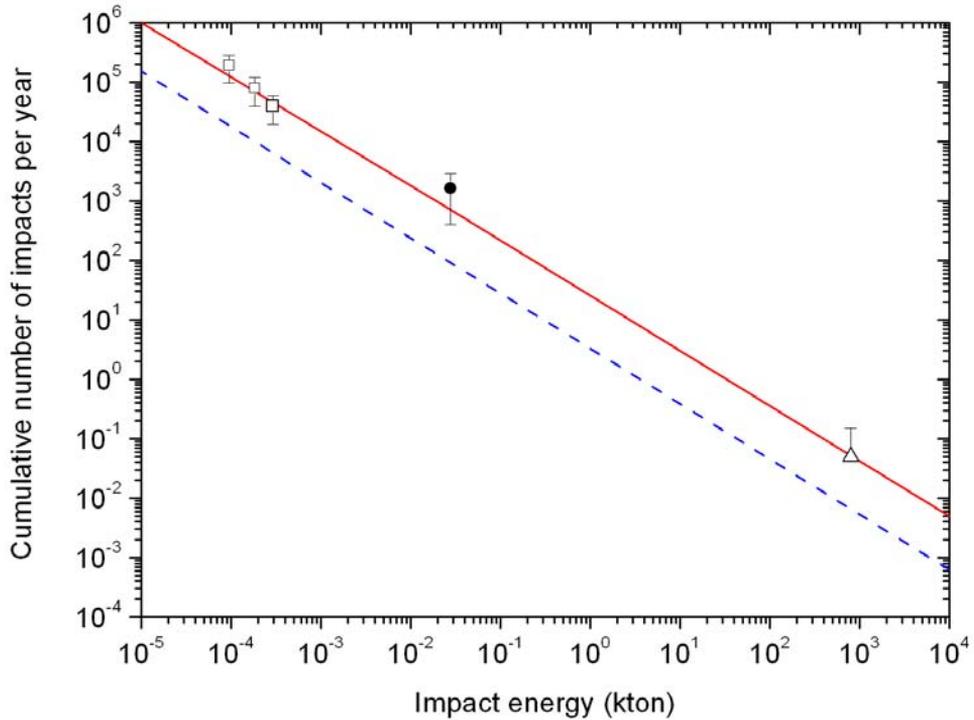

Figure 6. Cumulative number of impact events on Earth as a function of impact energy. The dashed line corresponds to the impact frequency derived by Brown et al. (2002). The squares correspond to the results derived from the lunar impact monitoring performed by Ortiz et al. (2006), while the solid line shows the frequency obtained by the same authors by assuming a luminous efficiency $\eta=2\cdot10^{-3}$. The result derived from the impact flash analyzed here is represented with a full black circle in this plot. The open

triangle corresponds to the flux derived by Brown et al. (2013) from the analysis of the Chelyabinsk event.